\documentclass[prc,final,5p,times,twocolumn,showpacs,showkeys]{revtex4-1}

\usepackage{lineno}
\usepackage{graphicx}
\usepackage{amssymb} 
\usepackage{url}

\newcommand{\UCN}{ultracold neutron (UCN)\renewcommand{\UCN}{UCN}}
\newcommand{\AMS}{accelerator mass spectroscopy (AMS) \renewcommand{\AMS}{AMS}}
\newcommand{\NIST}{the National Institute of Standards and Technology (NIST)\renewcommand{\NIST}{NIST}}
\newcommand{\MS}{Mass Spectrometer (MS)\renewcommand{\MS}{MS }}
\newcommand{\LANL}{Los Alamos National Laboratory (LANL)\renewcommand{\LANL}{LANL }}
\newcommand{\ATLAS}{Argonne Tandem Linear Accelerator System (ATLAS)\renewcommand{\ATLAS}{ATLAS}}
\newcommand{\ECR}{electron cyclotron resonance (ECR)\renewcommand{\ECR}{ECR}}
\newcommand{\SNS}{Spallation Neutron Source (SNS) at the Oak Ridge National Laboratory\renewcommand{\SNS}{SNS}}
\newcommand{\SPS}{Enge Split-Pole Spectrograph (SPS)\renewcommand{\SPS}{SPS}}
\newcommand{\RF}{radio-frequency (RF)\renewcommand{\RF}{RF}}

\begin{document}


\title{High-Sensitivity Measurement of $^3$He-$^4$He Isotopic Ratios for Ultracold Neutron Experiments }

\author{H.\,P.~Mumm}
\email{pieter.mumm@nist.gov}
\affiliation{National Institute of Standards and Technology, 100 Bureau Drive, Stop 8461, Gaithersburg, MD 20899, USA}
\author{W.~Bauder}
\affiliation{University of Notre Dame, 225 Nieuwland Science Hall, Notre Dame, IN 46556, USA}
\author{N.~Abrams}
\affiliation{Columbia University, 538 West 120th St., 704 Pupin Hall, MC 5255, New York, NY 10027, USA}
\author{M.\,G.~Huber}
\affiliation{National Institute of Standards and Technology, 100 Bureau Drive, Stop 8461, Gaithersburg, MD 20899, USA}
\author{C.\,M~Deibel}
\affiliation{Louisiana State University, 218-B Nicholson Hall, Tower Dr., Baton Rouge, LA 70803, US}
\author{C.\,R.~Huffer}
\affiliation{North Carolina State University, 2401 Stinson Drive, Box 8202, Raleigh, NC 27695, USA}
\affiliation{Triangle Universities Nuclear Laboratory, 116 Science Drive, Box 90308, Durham, NC 27708, USA}
\author{P.\,R.~Huffman}
\affiliation{North Carolina State University, 2401 Stinson Drive, Box 8202, Raleigh, NC 27695, USA}
\affiliation{Triangle Universities Nuclear Laboratory, 116 Science Drive, Box 90308, Durham, NC 27708, USA}
\author{R.~Janssens}
\affiliation{Argonne National Laboratory, Physics Division, Building 203, 9700 S. Cass Ave., Argonne, IL 60439, USA}
\author{C.\,L.~Jiang}
\affiliation{Argonne National Laboratory, Physics Division, Building 203, 9700 S. Cass Ave., Argonne, IL 60439, USA}
\author{K.\,W.~Schelhammer}
\affiliation{North Carolina State University, 2401 Stinson Drive, Box 8202, Raleigh, NC 27695, USA}
\affiliation{Triangle Universities Nuclear Laboratory, 116 Science Drive, Box 90308, Durham, NC 27708, USA}
\author{C.\,M.~Swank}
\affiliation{North Carolina State University, 2401 Stinson Drive, Box 8202, Raleigh, NC 27695, USA}
\affiliation{Triangle Universities Nuclear Laboratory, 116 Science Drive, Box 90308, Durham, NC 27708, USA}
\author{R.\,H.~Scott}
\affiliation{Argonne National Laboratory, Physics Division, Building 203, 9700 S. Cass Ave., Argonne, IL 60439, USA}
\author{C.\,M.~O'Shaughnessy}
\affiliation{University of North Carolina at Chapel Hill, 120 E. Cameron Ave., CB \#3255, Chapel Hill, NC 27599, USA}
\affiliation{Triangle Universities Nuclear Laboratory, 116 Science Drive, Box 90308, Durham, NC 27708, USA}
\author{R.\,C.~Pardo}
\affiliation{Argonne National Laboratory, Physics Division, Building 203, 9700 S. Cass Ave., Argonne, IL 60439, USA}
\author{M.~Paul}
\affiliation{Hebrew University of Jerusalem, Racah Institute of Physics, Jerusalem 91904, Israel}
\author{K.\,E.~Rehm}
\affiliation{Argonne National Laboratory, Physics Division, Building 203, 9700 S. Cass Ave., Argonne, IL 60439, USA}
\author{R.~Vondrasek}
\affiliation{Argonne National Laboratory, Physics Division, Building 203, 9700 S. Cass Ave., Argonne, IL 60439, USA}
\author{L.~Yang}
\affiliation{University of Illinois at Urbana-Champaign, 1110 West Green Street, Urbana, IL 61801, USA}
\date{\today}

\begin{abstract}
Research efforts ranging from studies of solid helium to searches for a neutron electric dipole moment require isotopically purified helium with a ratio of $^3$He to $^4$He at levels below that which can be measured using traditional mass spectroscopy techniques.  We demonstrate an approach to such a measurement using accelerator mass spectroscopy, reaching the $10^{-14}$ level of sensitivity, several orders of magnitude more sensitive than other techniques.  Measurements of $^3$He/$^4$He in samples relevant to the measurement of the neutron lifetime indicate the need for substantial corrections. We also argue that there is a clear path forward to sensitivity increases of at least another order of magnitude.  
\end{abstract}




\maketitle

\vskip-0.2in
\newpage

\section{Introduction}
\label{sec_intro}
Isotopically purified $^4$He is central to the success of a variety of experiments including the ultracold neutron lifetime measurement (UCN)~\cite{Seestrom2014,o2009measuring} at \NIST, torsion oscillator experiments studying solid $^3$He-$^4$He mixtures~\cite{PhysRevB.83.224519, Gumann2012b}, and the neutron Electric Dipole Moment (nEDM) experiment~\cite{Golub19941,nEDM2014} at the \SNS.   The common feature of these experiments is that each requires accurate measurements of the $^3$He-$^4$He ratio ($R_{34}$) at levels below that which can be measured using standard mass spectrometry techniques (the typical abundance sensitivity of a commercial mass spectrometer is $\approx10^{-9}$). 

In the neutron lifetime experiment, for example, it is essential to have significantly increased isotopic purity as the UCN loss rate due to the reaction $^3$He(n,p) is $r_{loss}=nR_{34} \sigma_{th} v_{th}$, where R$_{34}$ is the isotopic ratio of the helium, $n=2.17\times10^{22}$ cm$^{-3}$ is the number density of helium atoms at 300~mK, $\sigma_{th}$ = 5333 b is the $^3$He thermal neutron cross section, and $v_{th}$ = 2200 m/s is the thermal neutron velocity.  The current world average neutron lifetime is ($880.3\pm1.1$) s~\cite{PDG}, thus a purity of $R_{34} <5\times10^{-15}$ is required to reduce the fractional systematic correction to the neutron lifetime due to $^3$He to less than 0.1 s, the ultimate goal of the experiment. 

Secondly, there are puzzling observations in experiments designed to test the supersolid phenomenon in helium~\cite{Kim2004}. These observations are affected strongly by the solid $^4$He sample quality which depends on growth condition, sample geometry, and, importantly, the $^3$He concentration.  The surprisingly high sensitivity of the torsional oscillator (TO) frequency shifts to minute $^3$He concentrations at parts per million level has helped to explain the role $^3$He plays in these systems. Certain behaviors of these systems have been correlated with extrapolated concentrations down to $R_{34}\sim 10^{-14}$~\cite{Gumann2012b}.

Finally, in the case of the nEDM experiment, the neutron precession rate is measured for E and B fields parallel and anti-parallel. Because there are not enough neutrons to measure the precession signal directly, a spin-dependent nuclear interaction with polarized $^3$He is used.  This polarized $^3$He (in a vessel of superfluid $^4$He also containing the neutrons) is eventually depolarized by interactions with the container walls and must be removed from the system. 

 In order to do this, the heat-flush technique is used.  The heat-flush technique utilizes the fact that $^3$He atoms in He II form part of the normal fluid component. Thus in an apparatus that creates a thermal counterflow, that is, where normal fluid travels away from a heat source and superfluid simultaneously moves towards it, $^3$He atoms will tend to congregate at the cold end of the apparatus.  There appears to be no intrinsic limit to the isotopic purity that can be obtained with this method~\cite{Herzlinger197265, Atkins1976733, McClintock1978}. To validate heat-flush transport performed at the \SNS, measurements of test samples with concentrations ranging from about $10^{-8}$ to $10^{-12}$ are required.  The heat-flush technique is also used to isotopically purify the samples discussed  later in this paper.



Earlier work verifying the heat-flush has generally relied on methods of increasing the concentration of purified samples and then using traditional mass spectrometers.  This can be done, for example, by running samples through a purifier in reverse.   This approach has yielded indirect limits of $R_{34}< 5\times10^{-16}$~\cite{McClintock1978} using a one-shot purification and $R_{34}< 5\times10^{-13}$ for a continuous-flow apparatus~\cite{Hendry1987131} similar to that used to purify the helium used in this work.  There is no reason to expect that the purities obtained with the continuous-flow apparatus should be less, so it has been historically assumed that the purified helium used in the neutron lifetime experiment, which we will refer to as ultrapure, was $R_{34}\approx10^{-16}$; obviously a direct measurement is desirable.
Specialized commercial mass spectrometers can reach levels of sensitivity on the order of $1\times10^{-12}$.  
On the other hand, \AMS provides the only potential way to directly measure $R_{34}$ in isotopically purified helium samples at the level of sensitivity required for the neutron lifetime experiment and should be expected to reach an ultimate sensitivity at the $10^{-15}$ level.  A program to reach these levels of sensitivity was started in 2000.  Since then, significant progress has been made in developing the technique for AMS measurement of trace $^3$He impurities, and several purified samples of experimental significance have been successfully measured.  In this article we report the results of this effort.

\section{Experiment}
\AMS\ (see~\cite{Kutschera2005145} for a review) is a technique typically dedicated to the measurement of radionuclides of extremely low abundance, either of natural (cosmogenic or radiogenic nuclides) or artificial origin (produced via nuclear reactions).  The principle of the technique is based on the acceleration of ions of the specific nuclide at an energy sufficient for separation or discrimination from abundant isotopic and isobaric species and from stable molecular interferences of close-by mass. The high ion energy, compared to that used in standard mass spectrometric methods, provides unambiguous identification through a combination of magnetic and electrostatic analysis and nuclear detection methods (e.g. specific energy loss, ion range in matter, time of arrival).  In our case this separation was accomplished using a \SPS~\cite{Enge1979} as discussed below.
	
In the current project, the isotope to be detected is stable $^3$He whose isobar is radioactive $^3$H (t$_{1/2} =  12.33$ y), the only other A = 3 bound nuclide. There is a historical tie here, as $^3$He was the first nuclide to be identified after acceleration through a cyclotron by Alvarez and Cornog~\cite{Alvarez1939,Alvarez1981}, predating by about four decades the development of \AMS\ as a full-fledged technique. In our experiment, we take  advantage of the fact that acceleration of $^3$He to 3 MeV/nucleon allows us to dissociate and completely eliminate molecular species (such as H$_3^{+}$) likely to be present (the highest energy used in a run was 5 MeV/nucleon, but all data reported here used an energy of 3 MeV/nucleon). Although of different mass than $^3$He, this species could cause severe background in the case of detection of ultra-low $^3$He abundances. As described later, the difficulty in this measurement is not in identifying or detecting $^3$He but controlling the sources of atmospheric and laboratory background of helium with much larger $^3$He abundances than in the isotopically purified helium samples of interest.

\begin{figure*}
\centering
\includegraphics[clip=true, trim=00mm 8mm 00mm 00mm,width=0.85\textwidth]{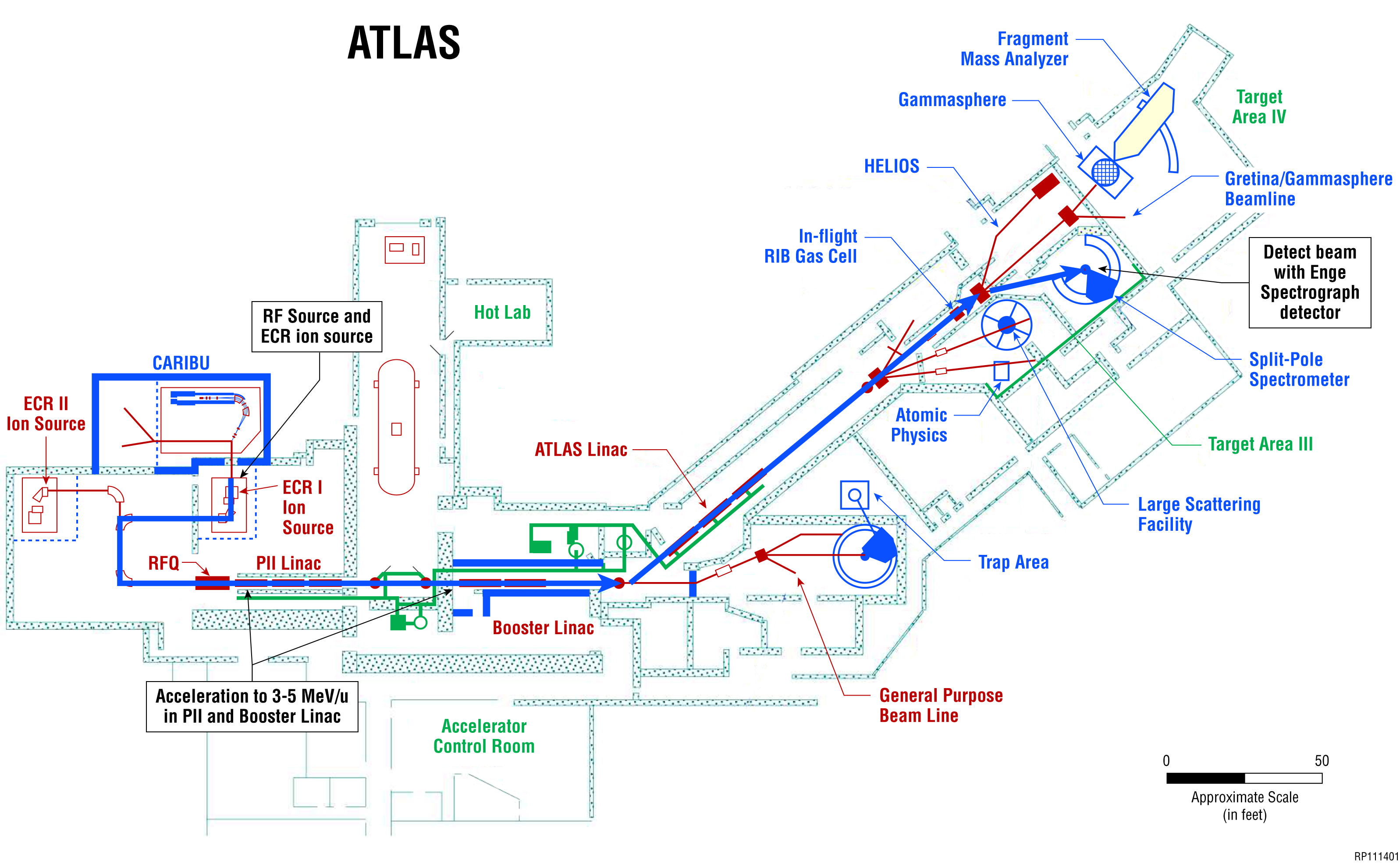}
\caption{Floor plan of the \ATLAS\ Accelerator Facility.  The ECR-1 ion source and an associated \RF\ discharge source were used for $^3$He \AMS\ and the beam was accelerated to 8~MeV in the ``PII" and ``Booster" linac sections.  The $^3$He produced was detected, after transport, in the Enge Split-Pole Spectrograph.}\label{fig:ATLAS_Layout}
\end{figure*}

Our effort was carried out using the \ATLAS\ at the Argonne National Laboratory.   \ATLAS~\cite{Bollinger1993221} is the world's first superconducting linear accelerator for heavy ions.  It consists of approximately 50 superconducting \RF\ resonators along with superconducting focusing solenoids and room-temperature beam transport elements designed to provide ions of any species at maximum energies from 25~MeV/nucleon for the lightest ions to 10~MeV/nucleon for the heaviest species such as uranium.  It is a national user facility for low-energy heavy-ion research.  Beam time at \ATLAS\ is in high demand, and it is difficult to plan experiments that require large quantities of beam time either for development or data collection.  Nevertheless \ATLAS\ has a long history of supporting development in accelerator mass spectroscopy.  In general, that effort has been focused on AMS for heavier isotopes that require the higher energy available from \ATLAS\ to allow unique discrimination of the isotope of interest from a stable isobar contaminant.  As noted, in this case we are using \AMS\ techniques to identify the level of concentration of $^3$He in $^4$He.  The use of positive ion sources and sophisticated detection systems made \ATLAS\ a good choice for the development of this technique.  The overall floor plan for \ATLAS\ is shown in Figure~\ref{fig:ATLAS_Layout} and the portions of the accelerator critical for $^3$He \AMS\ are noted by box call-outs.

Configuring the linear accelerator and beam transport system for an \AMS\ experiment requires establishing a hardware configuration that is set for an ion species with a specific mass to charge state (M/q $\approx 3$ for $^3$He$^+$) ratio and a specific initial velocity.  Thus the linac is typically tuned using a guide beam - a stable ion species that has a similar M/q ratio.  This tune is then scaled by a factor equal to the ratio of the exact value of M/q for the guide beam to that of the species of interest.  Here $^{12}$C$^{4+}$ was used for the initial accelerator and beam transport system setup.  The final beam energy used varied somewhat in successive experiments but was approximately 3~MeV per nucleon (MeV/nucleon).  That tune configuration was then scaled to the molecular species H$_3^{1+}$ as an additional check and to make the final tune to the beam cup located before the \SPS\ (see Fig.~\ref{fig:Source_layout}).  The scale factor from $^{12}$C$^{4+} \rightarrow $ H$_3^{1+}$ was 1.00783 and the final accelerator scaling from H$_3^{1+} \rightarrow$ $^3$He$^{1+} $was 0.99754.  The retractable beam cup can be replaced by a Au foil which strips accelerated $^3$He$^+$ ions to $^3$He$^{2+}$ and dissociates contaminant molecular ions (e.g. H$^{3+}$, DH$^+$)

\subsection{Description of Measurement Sequence}\label{section:Meas_intro}
\begin{figure}
\centering
\includegraphics[width=0.48 \textwidth]{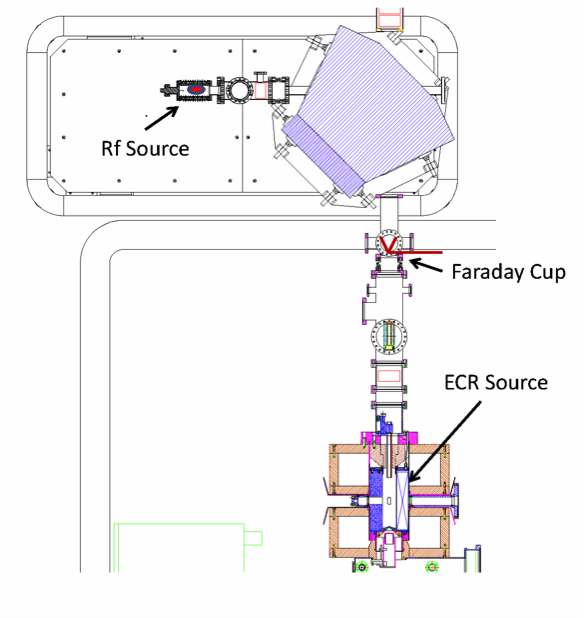}
\caption{Physical orientation of the \RF -discharge source, Faraday cup, and the ECR source used in this work.}
\label{fig:Source_layout}
\end{figure}

The \AMS\ $^3$He/$^4$He ratio is obtained by comparing the rate of detection of $^3$He$^{2+}$ ions in the spectrograph detector, corrected for detector efficiency and accelerator transmission, to the beam current of $^4$He out of the source.   The source operation was monitored by measuring the $^4$He$^{1+}$ current at the ion source Faraday cup shown in Figure~\ref{fig:Source_layout}.  This was accomplished by changing the ion source extraction voltage from 30~kV to 3/4 of 30~ kV (22.5~kV) to match the magnetic rigidity of the $^3$He$^{1+}$ ions.  Thus the source analyzing magnet was not changed during this cyclic process.  The beam is accelerated through ATLAS, passes through the gold stripper foil to remove unwanted molecules and to raise the $^3$He charge state to 2+, and then into the \SPS. The accelerator transmission was monitored periodically by returning the ion source to a hydrogen plasma and measuring the transmission to a Faraday cup at the position of the stripper foil of the SPS with a H$_3^{1+}$ beam.   Ions are detected in the SPS with an ionization chamber focal-plane detector that provides information both in position and in the  energy loss, $dE/dX$. This allows unique identification of $^3$He signals.

%

\subsection{Ion Source Development}
\label{sec:ion_source}

For the \ATLAS\ accelerator, the standard positive ion source is an \ECR\ ion source~\cite{Vondrasek2010}.  These sources have large vacuum chambers that have a significant internal surface area on which gases can be adsorbed.  For example, the \ECR\ source first used in this work had a 30 cm long, 8 cm diameter cylinder chamber.  In addition, such sources are often operated with helium as a support gas. The typical concentration of  $^3$He in atmospheric  $^4$He of $1.4\times10^{-6}$~\cite{Pobell1996} is to be compared to a sensitivity goal of one part $^3$He in  $10^{15}$ parts $^4$He.  This comparison provides a general perspective on controlling helium backgrounds.  It is readily apparent that it is not  possible to fully overcome helium outgassing from the cylinder walls (or possible atmospheric leaks) by simply operating the source at extremely high pressures and flow rates.

\begin{figure}
\centering
\includegraphics[width=0.45 \textwidth]{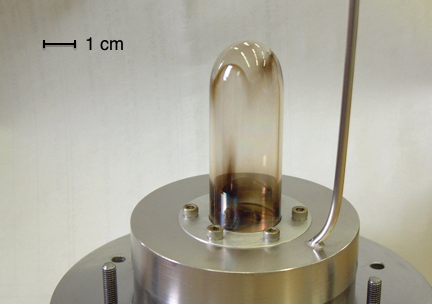}
\caption{The mini-\ECR\ source developed early in this work to reduce the $^3$He background seen in the main \ECR\ ion source.  The quartz tube is mounted on the extractor electrode for the \ECR\ source and extends into the \ECR\ cavity.  At the bottom of the quartz tube holding flange is the 1~mm extractor hole and the offset gas feed for the helium gas.}\label{fig:miniECR}
\end{figure}

To address this problem, a mini-\ECR\ ion source using the magnetic field and microwave power feeds of the existing source was developed.  The plasma volume was defined by a borosilicate glass tube attached to a redesigned aluminum extraction electrode system.  In principle this would separate the plasma from sources of natural helium background.  The mini-\ECR\ source designed for this experiment is shown in Figure~\ref{fig:miniECR}.

While this source geometry did allow measurements down to the $10^{-13}$ ratio regime or better, it had a number of drawbacks:
\begin{enumerate}
\item  The geometry did not allow the production of $^{12}$C$^{4+}$.  Therefore one had to return to the normal \ECR\ geometry for the guide beam and  re-install the mini-\ECR\ source after the linac was tuned.  Any problems that raised a question concerning the linac tune required removing this source and reverting to the standard \ECR\ geometry.  This cycling process was quite time-consuming.
\item Igniting the plasma in this geometry was occasionally difficult.
\item While the initial sensitivity observed for this geometry was much improved, it was still at least an order of magnitude above the most interesting regime of $10^{-14}$-$10^{-15}$.  
\end{enumerate}

\begin{figure}
\centering
\includegraphics[width=0.48 \textwidth]{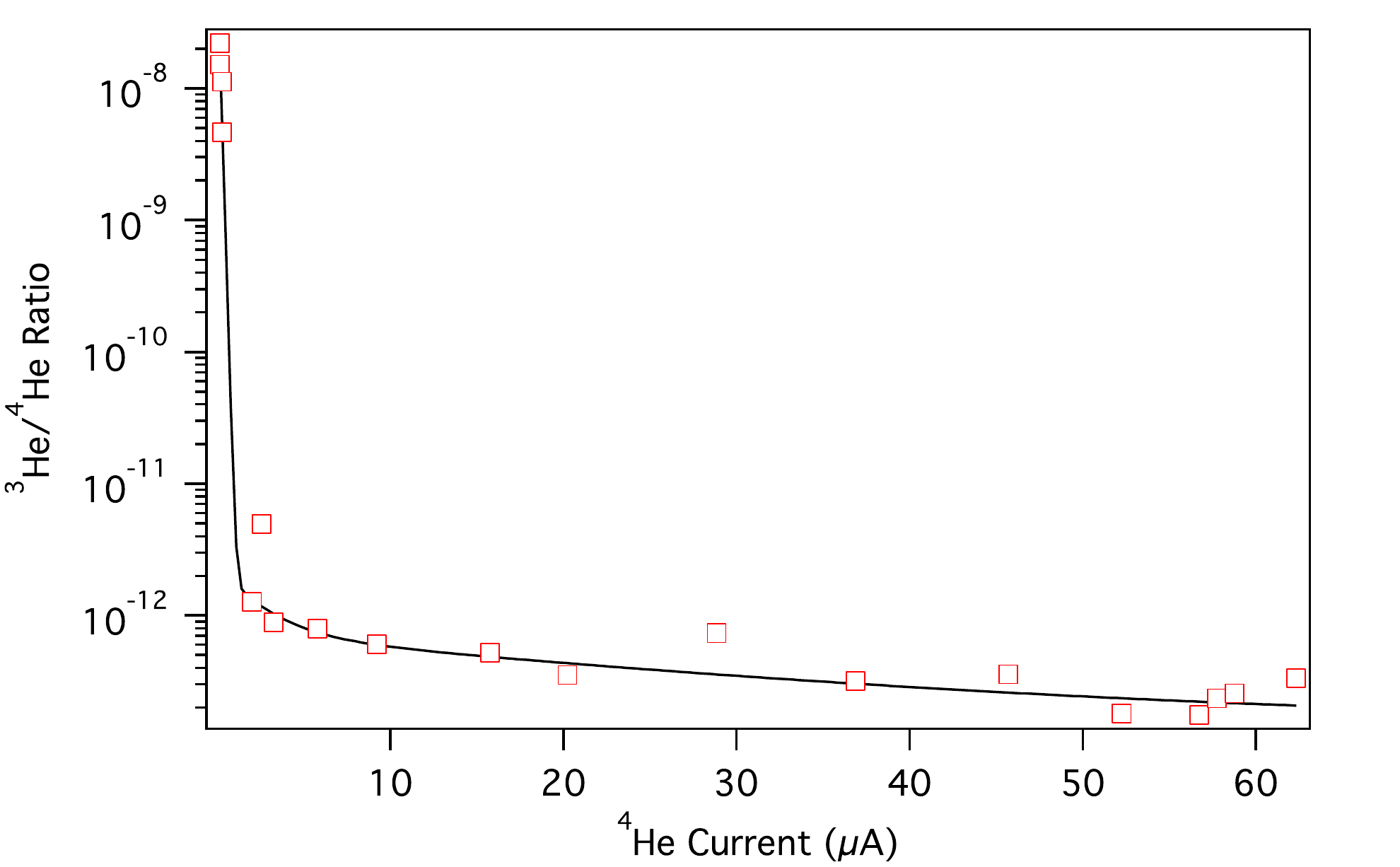}
\caption{The observed $^3$He/$^4$He ratio observed with an ultrapure sample as a function of the $^4$He beam current from the source.  The helium current is viewed as a surrogate to helium partial pressure in the source plasma.  The asymptotic value of $^3$He/$^4$He $\sim3\times10^{-13}$ observed at high $^4$He source current (for which source background becomes less significant) is consistent with that obtained using the current source configuration as described in Section~\ref{sec:final_results}. }\label{fig:3He4He_by_current}
\end{figure}

To illustrate these points, an example of a measurement sequence with this source configuration as a function of gas pressure is shown in Figure~\ref{fig:3He4He_by_current}.  This plot shows the $^3$He/$^4$He ratio starting out in the regime of natural abundance material with no flow of isotopically purified helium in the source. As a higher flow of purified gas is introduced into the source, the  $^3$He/$^4$He ratio approaches an asymptotic value that is interpreted as representing that of the actual sample. 

\begin{figure}
\centering
\includegraphics[width=0.49 \textwidth]{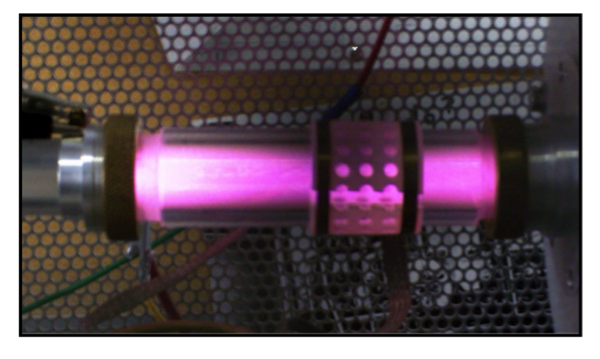}
\caption{\RF\ discharge source in operation with a helium plasma.  Gas flow from the helium sample enters from the left and ions are extracted though the electrode on the right.  \RF\ power is supplied using the coils visible to the center right of the source.}
\label{fig:Helum_Discharge}
\end{figure}

\begin{figure}
\centering
\includegraphics[width=0.48 \textwidth]{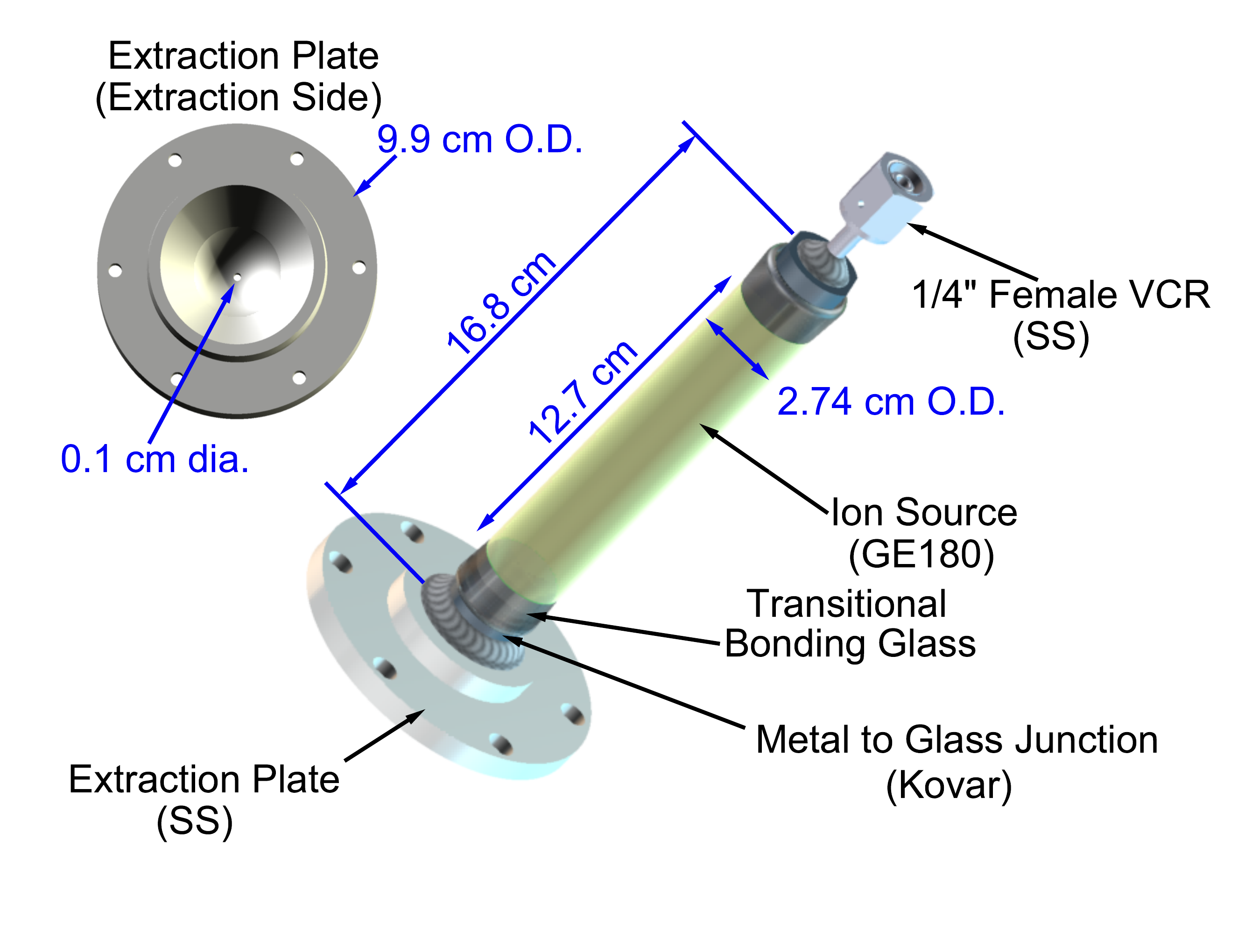}
\caption{The new \RF\  source plasma chamber constructed with materials having a low helium diffusivity.}\label{fig:RFSource_Chamber}
\end{figure}

Attempts were made to further reduce the source of helium background in this geometry by replacing the quartz (known to have a high helium diffusion constant) with a pyrolytic boron nitride tube.  These improvements achieved a sensitivity approaching $2\times10^{-14}$ in the  $^3$He/$^4$He ratio.  Nevertheless, the difficulty of working with the mini-\ECR\ source led us to explore other possibilities.  
The geometry for this two-source configuration is shown in Figure~\ref{fig:Source_layout}.  We adopted an \RF\ discharge ion source~\cite{Moak1951} that used a small quartz tube to define the plasma region and a simple inductive coupling scheme for \RF\ power into the source.  By utilizing a source completely separate from the \ECR\ source, we made it possible to quickly switch between the beams of interest ($^3$He$^{1+}$, H$_3^{1+}$, and $^{12}$C$^{4+}$) to check beams tunes and to make quick repairs and modifications to the source.
Figure~\ref{fig:Helum_Discharge} shows the \RF\ discharge source in operation with a helium plasma and Figure~\ref{fig:RFSource_Chamber} shows a more detailed drawing of the source glass and vacuum definition.

The \ECR\ remained the ion source for the $^{12}$C$^{4+}$ guide beam used to determine the initial accelerator  configuration.  The \ECR\ source plasma is then turned off and gas flow into the \ECR\ source stopped.  The bending magnet between the RF source and the ECR source is set to transport a specific ion from the \RF\ source through the \ECR\ and out into the connecting low-energy beam transport (LEBT) system.  

A molecular beam of H$_3^{1+}$ is used to retune the accelerator system from the ion source to the detection system, based on the previous $^{12}$C$^{4+}$ tune.  This beam is created by the \RF\ source using a hydrogen plasma with a current of 10-20 nanoamperes~(nA), allowing us to measure the accelerator beam transmission.  Additionally, we note that the ability to operate the source with only pure hydrogen allowed a measurement of the residual helium in the system and established a baseline for $^3$He background as discussed in Section~\ref{sec:final_results}.

To further reduce natural helium backgrounds a new plasma chamber for the \RF\ discharge source was designed.  The plasma chamber was intended to be easily swappable, allowing transmission measurements with a calibrated ($\approx$1\% natural) helium sample to be followed by ultra-pure measurements using an uncontaminated chamber to minimize the creation of large backgrounds for the isotopically purified samples.  The new plasma chamber is 16.84~cm long by 2.74~cm in diameter and constructed using materials with inherently low helium content, a reduced probability of adsorption of helium onto the surface, and low helium diffusivity. The new chamber shown in Figure~\ref{fig:RFSource_Chamber} is constructed from Kovar~\cite{disclaimer} bonded directly to Corning 7056, a borosilicate glass, and GE180, an aluminosilicate glass, with helium permeation rates several orders of magnitude lower than quartz.  GE180 has been used very successfully in polarized $^3$He neutron spin filters for many years~\cite{Chen2011}.

\subsection{Source gas handling system}\label{sec:gas_handling}
Ê
As the experiment requires a stable and well-defined accelerator tune, it is important to have the capability to switch between samples and plasma conditions with as little impact on the tune as possible.  For this to be true, the ion source operation must also be extremely stable and reproducible. In addition, the experiment requires the ability to compare samples with different isotopic ratios.

A gas handling system was built that allowed the switching of gas samples without turning off source voltages and without stopping the plasma operation.  This allowed frequent returns to an ultrapure hydrogen plasma in order to check the beam tune. In addition, using the same gas handling system allowed for remote switching between an ultra-pure sample and a reference sample, again while the voltages remained on.

\begin{figure}
\centering
\includegraphics[width=0.48 \textwidth]{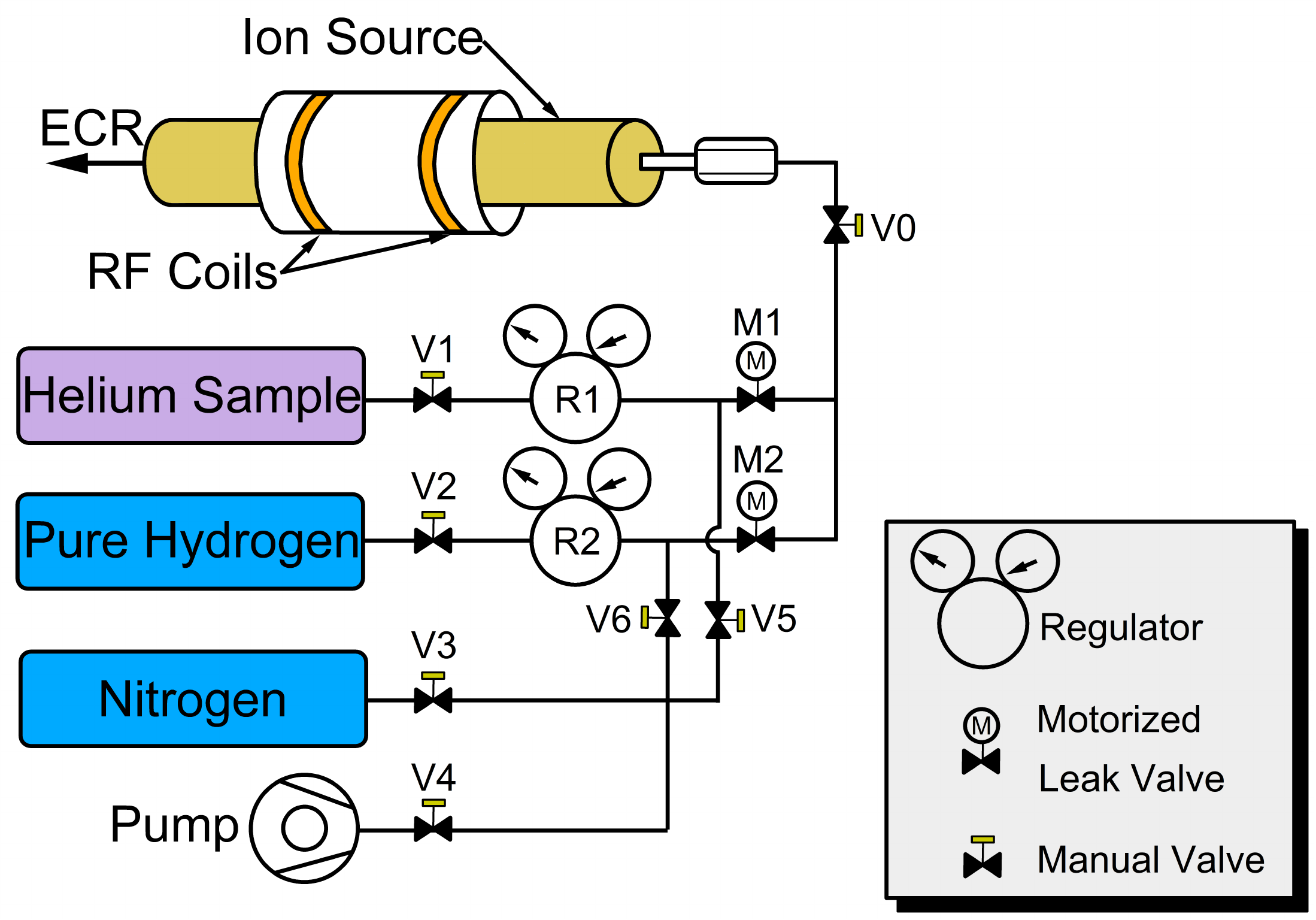}
\caption{Schematic of the gas handling system used with the \RF\ discharge source.  Two computer-controlled leak vales allow simple switching between samples while the source remains in operation. The nitrogen supply and pump were used to pump and purge the manifold assembly.}\label{fig:bad_sch}
\end{figure}

The gas handling system is shown in Figure \ref{fig:bad_sch}.Ê
This setup is rather typical of ion source manifolds but is completely new and has never been exposed to natural helium.  In addition, material choices for the manifold were made to minimize those that would easily absorb gasses, particularly helium, and thus lead to the potential contamination of  high-purity samples.  Central to the gas handling system design are two precision leak valves (Agilent Technologies model 951-5106).ÊÊ Each  leak valve controls the gas flow from a separate gas bottle.Ê The two leak valves were remotely operated by a DC servo motor located outside of the \RF\ discharge source.Ê This allowed the safe and controlled injection of gas into the ion source while lit.Ê
To avoid the possibility of helium diffusion into the system, all joints in the gas handling system were constructed of welded stainless steel or metal-metal compression fittings.  The entire gas handling system was baked out and leak tested at NIST with hydrogen prior to use at ATLAS.Ê Once installed at ATLAS the system was purged with dry boil-off nitrogen and pumped out.  Because of the low operating pressure of the source, an absolute pressure regulator (Airgas Y11 C440N) was mounted on the helium sample bottle.
Ê

\subsection{Detection Setup} The accelerated $^{3}$He$^{1+}$ beam was delivered after acceleration to the \ATLAS\ split-pole spectrograph where it was stripped to $^{3}$He$^{2+}$ by a 300 $\mu$g/cm$^2$ gold foil.  In addition the stripping process dissociated any contaminant molecular species such as residual molecules from the reference H$_3^{1+}$. 
This realtively pure $^{3}$He$^{2+}$ beam is then bent by the SPS into the focal plane detector.  As a final filter, the spectrograph completely separates any remaining  molecular fragments, disperses the $^3$He ions by momentum and focuses them onto the focal plane. The ions are then detected and counted by a parallel-grid avalanche counter (PGAC) followed by a multi-anode ionization chamber~\cite{Paul1997}. The PGAC provides x-position (horizontal dispersion plane) and  y-position (vertical plane) signals for individual ions and one of the anodes (dE4) of the ionization chamber provides an energy-loss signal distinctive for  $^{3}$He$^{2+}$  ions.

\section{Sample Preparation}
Validation of the the mass-spectroscopic measurements relied on comparison to samples prepared to have a well-known isotopic ratio.  In addition, samples of highly purified hydrogen gas were used to operate the source without helium and to provide a way of measuring the inherent $^3$He background of the \RF\ source and accelerator.
\subsection{Ultra-pure Hydrogen sample}
\label{ultrapure_hydrogen}
A commercial electrolysis hydrogen generator (Parker Balston H2-1200) with a palladium membrane was used to prepare samples of hydrogen gas with a very low helium content.  The generator was leak-checked by the manufacturer with one bar of helium and no observable leakage through the membrane was detected.  Assuming a natural helium isotopic ratio, this allows the maximum $^3$He  mass transfer rate through the membrane at this pressure differential to be estimated at $3\times10^{-20}$ mol/s.  The generator produces H$_2$ gas at $1\times10^{-6}$ mol/s.  Assuming that natural helium had diffused into the water used in the generator and reached equilibrium ($2\times10^{-8}$ mol$_{^3He}$/mol$_{H_2O}$) and that no helium remained in the all-metal sample bottle after evacuation and baking, we can place an upper limit on the possible $^3$He contamination during normal operation of less than $1\times10^{-21}$.  This should be understood to be a very rough estimate.  Nonetheless, in the context of the work reported here, the H$_2$ sample is free of $^3$He.

\subsection{Reference sample preparation}
\label{sec:reference_samples}
	 Two reference samples,  1 and 2, of well-known concentration were prepared by mixing ultrapure helium purified by the McClintock group using the heat-flush technique~\cite{McClintock1978} with natural helium from a bottle of 99.999\% commercial BIP (Built-In Purifier) helium.  At the onset of this work the ultrapure helium was expected to be $R_{34}\approx10^{-16}$ as described above, however in the mixing calculations that follow, the value of $R_{34}=(1.2\pm0.4)\times10^{-12}$ eventually measured for the ultrapure gas (sample 4, see Table~\ref{tbl:result}) was used, and results in a slight upward shift in concentration (see Section~\ref{sec:final_results} for a description of this measurement).  The $^3$He/$^4$He ratio of the commercial BIP helium was measured to approximately 1\% precision using traditional mass spectroscopy~\cite{OregonLab}.
Measurements of each BIP sample were bracketed with gas containing no helium to eliminate backgrounds and cross-contamination. Uncertainties are statistically dominated, thus we combine the standard uncertainties and obtain a result of R$_{34}$(BIP)$=(2.201\pm 0.005)\times 10^{-7}$.  This is consistent with a radiogenic helium signature, as would be expected from commercial compressed gas since such helium comes from crustal helium in natural gas wells~\cite{Pobell1996}. 

The mixing apparatus consisted of an assembly of 1/4-inch Swagelok Variable Compression Ratio (VCR) components that allowed connections for a bottle of the commercial helium, a bottle of purified $^{4}$He, a well-determined evacuated expansion volume, a precision pressure gauge (Paro-scientific model 745-400), a pump-out port, and a series of all-metal valves that allowed for control of gas flow between each of these components.  The stainless-steel sample bottles and expansion volume had fully welded fittings and all-metal sealed bellows and metal-seated valves.  They were pumped under vacuum and heated to a modest 46 C for three hours prior to use.   Measurements of pressure changes with the precision pressure gauge were used to determine all of  the volumes in the system.  

The mixing process used the following sequence.  A 300~ml stainless-steel sample bottle filled with ultra-pure helium was attached to the gas handling system.  Natural (BIP) helium was then allowed to fill the tubing (mixing volume) connecting the various valves (approximate volume 14~ml).  The natural gas bottle was valved off, and the gas was allowed to expand into an evacuated 150~ml volume, reducing its pressure by roughly a factor of ten.  Finally this bottle was valved off, and the residual gas allowed to mix with the ultra-pure sample for 30-45~min.  This process was repeated, giving a dilution factor of approximately 200 at each stage and yielding a concentration of approximately $1\times10^{-9}$ and then $5\times10^{-12}$ for samples 1 and 2 respectively.  The  pressure gauge  was used to record the pressures throughout the mixing process allowing precise calculations of the concentrations to be made.   

\subsubsection{Mixing simulations}

A mixing time of 30-45 minutes is too short to completely reach equilibrium.  To calculate the concentration of the reference samples, a detailed finite element analysis simulation of the gas-mixing system was performed.  The calculation included an approximate geometrical model of the system and the pressures of each sample as a function of time.  Temperatures were not recorded but are believed to have been relatively constant. 

The ratio of the sample volume to the mixing volume was determined by measuring the pressure change of the sample bottle when opened to the evacuated mixing volume.  Four consistent expansion measurements were made, two with the 300~ml gas cylinder and two with the 150~ml expansion cylinder.  If these volume measurements are averaged we obtain $(14.5 \pm 0.5)$ ml for the mixing volume, where the uncertainty is dominated by the manufacture's quoted uncertainty in the cylinder volume (10\%; three different cylinders).  A fifth ratio measurement was made using a more complicated sequence of pressure comparisons.  This measurement yields $(12.2 \pm 1.4)$~ml.   The uncertainty in this measurement is correlated with two of the other measurements.  We  take the simplest pressure based ratio measurements as the best value, but expand the uncertainty to account for the differences between the three methods, giving $(14.5 \pm 1.5)$ ml.   This is consistent with the volume calculated based on engineering drawings of the individual components.  The uncertainty in the volumes was simulated by adjusting the volume of the valve interiors within the model.  

While the initial conditions of the system prior to mixing are well known, the exact response of the system to the opening of the valve was not modeled.  To account for this, two limiting cases were considered.  First, the gas in the mixing volume was assumed to be completely mixed during the turbulent flow associated with opening the valve between the low-concentration, high-pressure (0.28 MPa) sample and high-concentration, low-pressure (0.021 MPa) sample, and second, the gas was assumed to be unmixed, with the low-concentration gas compressed into the remaining three valves by the incoming higher-pressure gas and the subsequent mixing through diffusion.   These limiting cases predict up to a 10\% difference in final concentration.  The uncertainty in the diffusion constant of $^3$He in $^4$He is roughly 4\%~\cite{Marrero1972}.  We treat this as independent of concentration.

The initial concentration of the ultra-pure, sample 4, is assumed to be that measured at Argonne as described in Section~\ref{sec:final_results}.  The calculated concentrations and associated uncertainties, including the propagation of the uncertainty in the measured concentration of sample 4, are shown in Table~\ref{table:MixingResults}.  

\begin{table}[!htb]
    \begin{tabular}{l|l|l} 
    &Sample 1&Sample 2\\
    \hline
      Configuration & R$_{34}(10^{-10})$ & R$_{34}(10^{-12})$\\
 \hline
      14.5~ml, mixed & $6.07\pm0.28$ & $2.62\pm0.07$\\
      14.5~ml, unmixed & $5.46\pm0.23$&$2.56\pm0.06$ \\
      12.7~ml, mixed& $5.79\pm0.23$&$2.49\pm0.10$\\
      12.7~ml, unmixed& $5.34\pm0.21$&$2.63\pm0.11$\\
\hline
    \end{tabular}
  \caption{Summary of results from the concentration calculations for various configurations.  The nominal configuration is 14.5~ml unmixed.  12.7~ml was used in place of 13~ml to estimate the effect of 1 standard error in volume before the final uncertainties were determined for this work.  Uncertainties are combined standard errors (1 $\sigma$).  Calculations are compared to measurement in Figure~\ref{fig:summary}.}
  \label{table:MixingResults}
\end{table}

The uncertainty in volume results in a roughly 4.5\% and 4.9\% change in concentration for the first and second stages of the mixing process respectively (in the mixed case). The uncertainty due to a lack of full knowledge of how the gas mixes is between 5\% to 10\% while the uncertainty due to the diffusion coefficient is  4\%. Assuming that the uncertainties can be added in quadrature, the total uncertainty is then 9.5\% for the first stage and 13\% for the second, yielding concentrations of R$_{34}=(6.1\pm0.6)\times10^{-10}$ and R$_{34}=(2.6\pm0.3)\times10^{-12}$ for sample 1 and sample 2 respectively.  These values include the 8.4\% uncertainty in the measured ultrapure sample used in the mixing.


\section{Comparative Results}
While it is possible to report an absolute isotopic ratio, a  potentially more precise approach is to make a comparative measurement  between the samples of interest and the well known reference samples described in Section~\ref{sec:reference_samples}.  In addition, we report results carried out with less sensitive traditional mass spectroscopy. 

\subsection{Oregon}
	 The two reference helium samples (1 and 2) were measured with a sensitivity of roughly $1\times10^{-9}$ using traditional mass spectroscopy at the Helium Isotope Laboratory~\cite{OregonLab}.  As with the BIP helium, sample measurements were bracketed with gas samples expected to contain no $^3$He to eliminate backgrounds and cross-contamination. Uncertainties are again statistically dominated.   The samples were found to have isotopic ratios of $(1.93\pm0.54)\times 10^{-9}  $ and $(5.2\pm4.7)\times 10^{-10}$ for  1 and 2, respectively (standard errors indicated).  
	 
%
	
\subsection{ATLAS}\label{sec:final_results}
\AMS\ measurements were performed at \ATLAS\ on four separate samples: the two reference samples, an ultrapure helium sample that was extracted from the original shipping cylinder and hence was expected to be uncontaminated, and a sample of helium extracted from the UCN lifetime apparatus described in Section~\ref{sec_intro}.  As noted in Section~\ref{section:Meas_intro} the $^{4}$He and $^{3}$He currents differ by several orders of magnitude requiring that different detection methods are used for counting the two species.  Thus, to measure the isotopic ratio, $^{3}$He counts on the SPS focal plane detector are compared with $^{4}$He beam current measured out of the ion source.  This method relies on knowledge of the transmission efficiency through the accelerator and detector system.  The $^{3}$He/$^{4}$He ratio can be expressed as

\begin{equation}
R_{34} = \frac{N_{3}}{I_{4}\epsilon_{SPS}Tt},
\label{Eqn:ratio}
\end{equation}

\noindent where $N_{3}$ are the background subtracted counts of $^{3}$He in the detector, $I_{4}$ is the $^{4}$He beam intensity at the ion source (measured in ions s$^{-1}$), $t$ is the $^{3}$He counting time, $T$ is the accelerator transmission, and $\epsilon_{SPS}$ is the detector efficiency.  The accelerator transmission from the ion source Faraday cup to the spectrograph stripper foil is measured using the molecular hydrogen beam, H$_3^{1+}$.  To alleviate any  concerns over the reproducibility of scaling between tunes, the primary samples were measured back-to-back, only measuring the transmission before and after the sequence.  The drift in the accelerator transmission is assumed to be linear and the average value of these transmission measurements was used to calculate $R_{34}$.  For the data presented here, the average transmission was 20\%.  After the run it was discovered that a magnet on the low energy beamline was not set with sufficient precision to ensure reproducible switching between $^3$He$^{1+}$ and H$_3^{1+}$.  This introduces a potentially large uncertainty  into the determination of the transmission because the H$_{3}$ transmission measurements may not accurately reflect the $^{3}$He transmission if this magnet was not precisely scaled.  Off-line tests have shown that this effect introduces a 24\% uncertainty in the transmission.  Fortunately, all the samples were measured in a continuous sequence without adjusting the magnets or accelerator parameters, meaning that this uncertainty can be considered a systematic that  effects all the sample measurements equally.  Nonetheless, we conservatively assign a 24\% uncertainty to the transmission measurement.

In order to normalize $^{3}$He counts to $^{4}$He output, the $^{4}$He current was periodically measured by scaling the injection voltage out of the source and reading the current on the Faraday cup after the first dipole magnet as shown in Figure~\ref{fig:Source_layout}.  Typically the $^4$He current was between 360~nA to 400~nA.  The source stability and performance was checked by measuring the $^4$He current before and after each sample.  $^3$He runs are typically one hour  and the measured $^4$He current before and after each run was averaged for use in Equation~\ref{Eqn:ratio}.  The variation in $^4$He current did not exceed 10\% during any data run.

\begin{figure}[h!]
\begin{center}
\includegraphics[width=9cm]{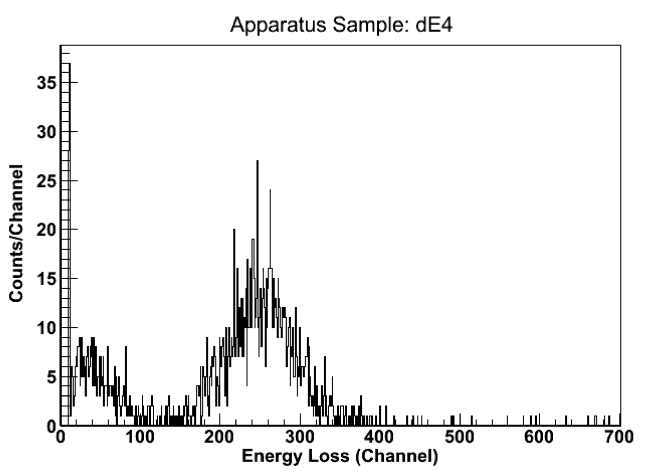}
\caption{$^3$He energy loss in anode 4, dE4, of the focal plane detector.   Events are required to be between channels 175 and 350.}{\label{fig:He3app} }
\end{center}
\end{figure}

On the focal plane of the spectrograph, the particles are detected with a combination of gas detectors.  The PGAC measures position on the focal plane and an ionization chamber measures the energy deposition across 5 different anodes.  Because the stripping foil provides excellent rejection of  H$_{3}^+$,  position information was unnecessary to distinguish it from $^{3}$He$^{2+}$.  $^{3}$He ions were instead identified through cuts on energy deposition.  Figure \ref{fig:He3app} shows a typical plot of the energy loss in anode 4, dE4.  dE4 was used for $^{3}$He identification because it showed the greatest separation of the $^{3}$He from the noise in the lower channels. 

The energy spectrum has a distinct double peak structure.   The larger peak is identified as $^{3}$He while the smaller peak at low channel number is a combination of  energy-degraded $^{3}$He, effects from cosmic rays, and detector noise.   $^{3}$He events were defined by a cut which was placed around the larger peak (channels 175-350).  The  detection efficiency for $^{3}$He produced by this cut combined with effects of the \SPS\ geometry was measured using the 8.78 MeV alpha from the $^{212}$Po daughter of a calibrated $^{228}$Th source placed at the object position of the spectrograph (the same position as the stripper foil). Using only the dE4 signal the efficiency was found to be $(70\pm7)$\%. The  30\% losses are estimated to be: 10\% blockage through wire planes and 20\% due to inefficient triggers in the PGAC and largely constitute the peak seen in Figure~\ref{fig:He3app} at low channel number. 


\begin{table}[h!]
\begin{center}
\begin{tabular}{ l || c | c}
Sample Number & R$_{34}$ & Uncertainty \\
\hline
1 (reference) & 3.8 x 10$^{-10}$ & 1.0 x 10$^{-10}$\\
2 (reference)  & 3.0 x 10$^{-12}$ & 0.8 x 10$^{-12}$\\
3 (apparatus) &  1.8 x 10$^{-12}$ & 0.6 x 10$^{-12}$ \\
4 (ultrapure) &  1.2 x 10$^{-12}$ & 0.4 x 10$^{-12}$\\
\hline 

\end{tabular}
\caption{\AMS\ determined isotopic ratios of the four samples described in the text; the two samples prepared as known references, the sample extracted from the neutron lifetime apparatus, and a sample representing the original source of ultrapure. }
{\label{tbl:result} }
\end{center}
\end{table}

The $^{3}$He concentration in each sample was measured in three sequential runs each with sufficient statistics to make the statistical errors negligible.  R$_{34}$ was calculated as the unweighted mean of these three runs. The uncertainties for each set of measurements are dominated by the systematics due to small fluctuations in the ion source output and accelerator transmission that cannot be tracked in real time.  The uncertainty due to these fluctuations is quantified by the standard deviation of the three measurements, assuming a normal distribution.  This uncertainty is around 10\% for each sample.

\begin{figure}[htb!pb]
\centering
\includegraphics[width=0.5 \textwidth]{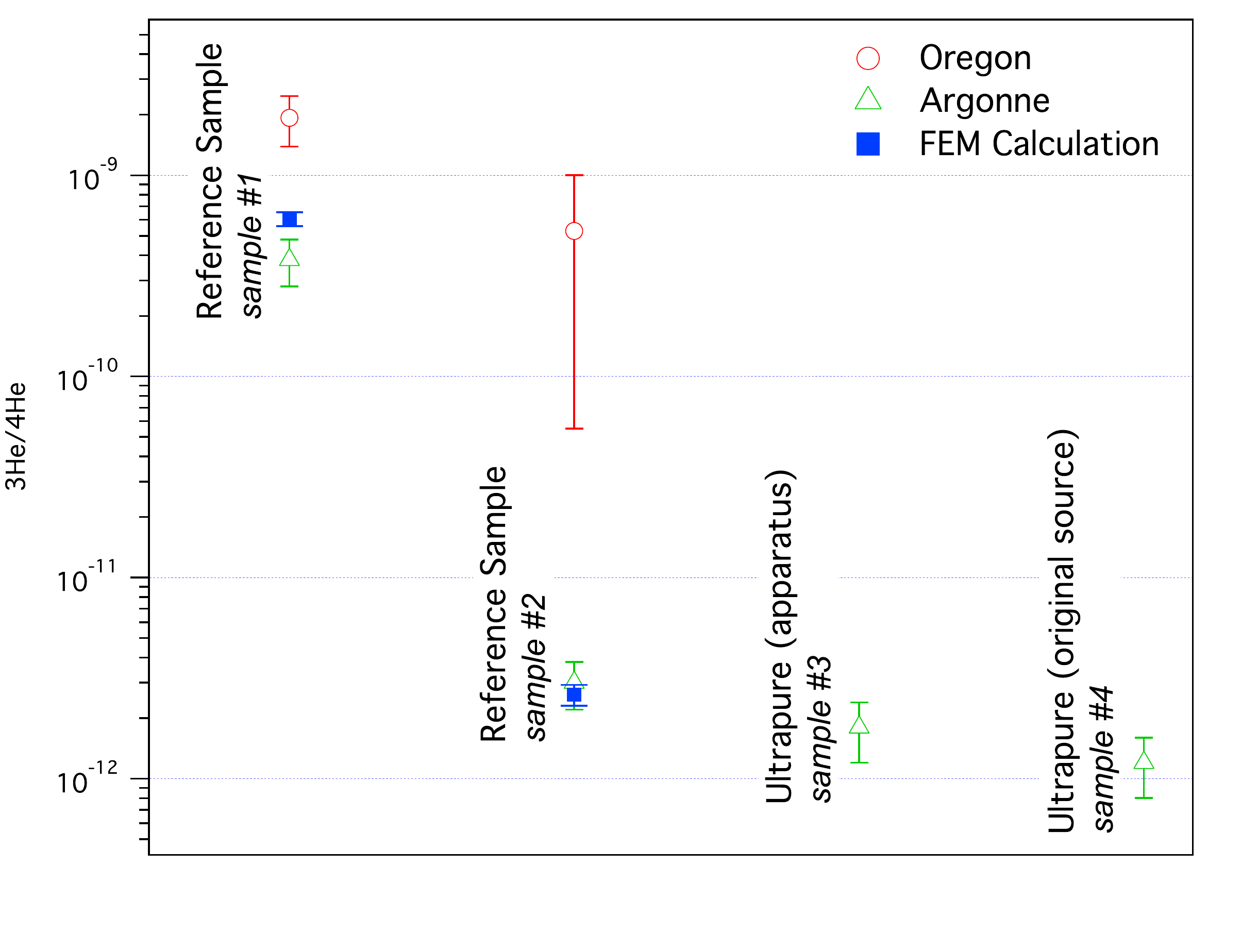}
\caption[]{Comparison of results, good agreement is seen between the Argonne measurements, $\triangle$, and the calculated concentration of the reference samples, $\blacksquare$.  Error bars show the combined standard uncertainty.}
\label{fig:summary}
\end{figure}

We combine all uncertainties in quadrature.  The results are tabulated in Table~\ref{tbl:result}.  As seen in Figure~\ref{fig:summary}, good agreement is seen between the calculated concentrations of the prepared reference samples and the results of AMS.  Importantly, the ultrapure samples extracted from the UCN lifetime apparatus show much higher concentrations of $^3$He than was assumed in planning the experiment, although difficulties with the extraction procedure leads to some question as to how well this sample represents the concentration as seen by the trapped neutrons during data collection.  Nonetheless a concentration of $R_{34}=(1.8\pm0.6)\times10^{-12}$ results in a systematic shift in the measured lifetime of $-(34.1\pm11)$~s, much larger than the ultimate goal of the measurement.  We note also, that $^3$He was measured in sample 4.  It is not clear when this contamination occurred, or whether it has implications for the purification process.  To address these questions we have built a new purifier and are preparing a new set of measurements that will be the subject of a future publication.   Finally, we note that improvements in the \AMS\ measurements are are expected. For example, a more precise Hall probe has been installed in the RF ion source injection magnet, and in future experiments, the setting of this magnet should be reliable.  In addition, a different choice of detector, for example a solid-state detector, is expected to reduce the uncertainty in detector efficiency.

\begin{figure}[h!]
\centering
\includegraphics[width=0.5 \textwidth]{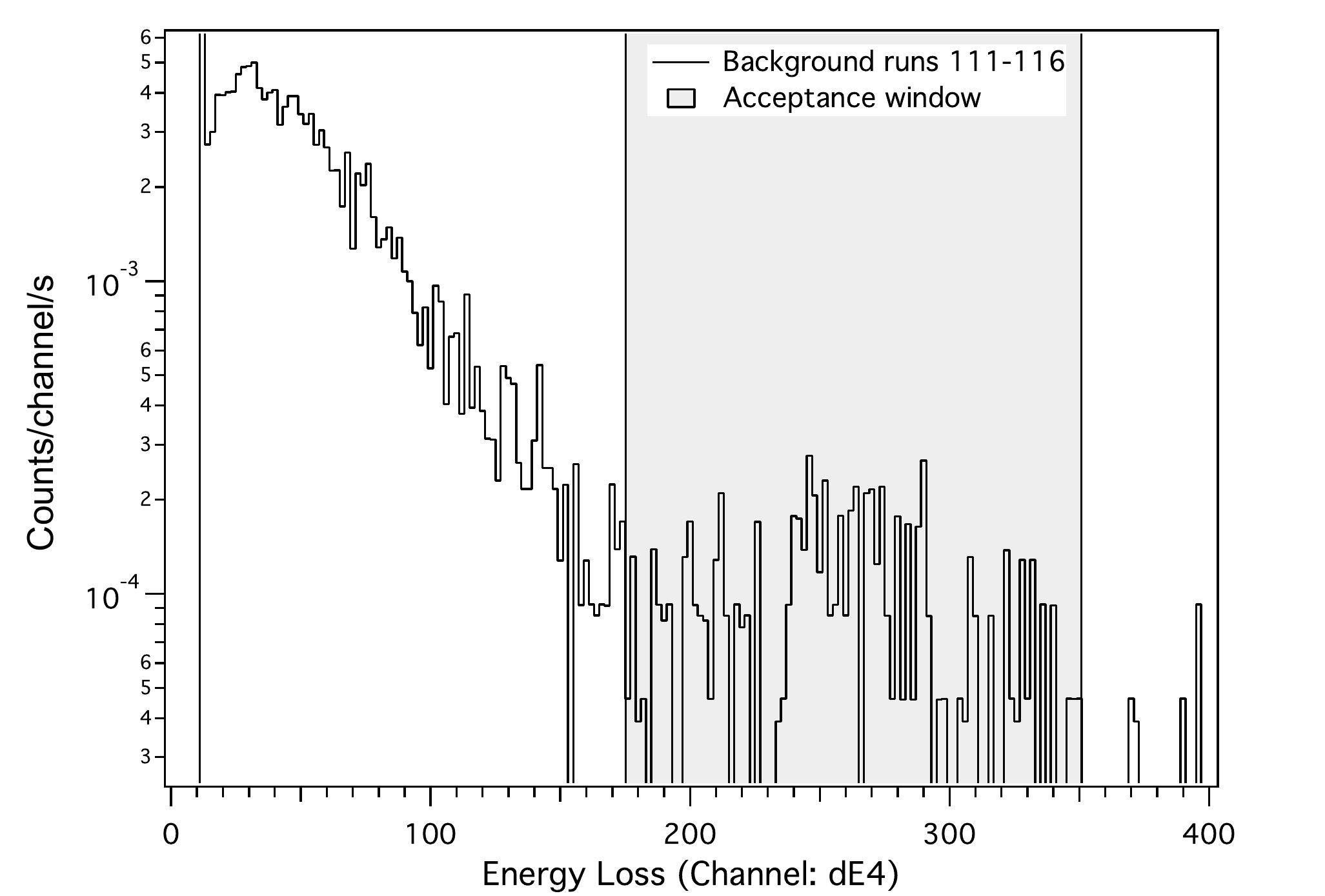}
\caption[]{Count rate in spectrograph after running pure hydrogen in the source for roughly 45 minutes. Natural $^3$He backgrounds fall in the region from 175 to 350 in dE4}
\label{fig:summary2}
\end{figure}

Near the end of the measurement, a series of runs were carried out with the ultrapure hydrogen described in Section~\ref{ultrapure_hydrogen} in the source.  These runs followed measurements taken with the $\approx 10^{-9}$ reference sample, and were intended to determine the time constant of residual helium gas in the system.   The count rates were seen to fall fairly rapidly to a stable rate with a time constant of $\approx$ 730~s.  From Figure~\ref{fig:summary2}, the constant background rate is seen to be $7.6\times10^{-3}$~s$^{-1}$.  Assuming the behavior of the source is the same when running hydrogen and helium, the limiting sensitivity from this constant background can be calculated to be between $2\times10^{-14}$ and $3\times10^{-14}$.  However, as described in Section~\ref{sec:gas_handling} both sides of the gas handling system are connected directly to the \RF\ source chamber at all times.  Thus we believe that the constant $^3$He background is likely the result of the valves not closing completely, and  does not represent a background for the helium sample runs.  It will be possible to verify this in future experiments by evacuation and then back-flushing both sides with ultrapure hydrogen.

\section{Conclusions and Future Work}

We have used \AMS\ to perform absolute measurements the isotopic ratio of $^3$He to $^4$He at the level of $10^{-13}$ and validated these measurements by comparison to known reference samples.  The absolute ratios show good agreement with our produced standard concentrations.  In addition we have shown that natural helium backgrounds can be controlled to at least the $10^{-14}$ level.  Furthermore, we have reason to believe that the contamination at this level was due to a leaky valve, and therefore expect that background can be reduced at least an order of magnitude.  Systematic problems with the measurement, in particular setting and measuring the magnetic field of the steering magnet in the low energy beamline have been solved or, in the case of the detector efficiency, replacement of the gas counter by a solid-state detector would resolve the issue.  We believe that it should be possible to reach measurement sensitivities on the order of $10^{-15}$ at which point statistics will become a limiting factor (for context, the $^3$He count rate for the $\sim 10^{-12}$ samples was less than 1 s$^{-1}$.   This level of sensitivity would be sufficient to allow future versions of the neutron lifetime measurement to reach the 0.1~s level.  Perhaps the most important result of this work is the demonstration that the $^3$He/$^4$He concentration of the ultrapure gas sample, originally expected to be in the range of $10^{-16}$ based on the heat flush purification technique, was in fact  $\sim1\times10^{-12}$. This is critical in understanding the large systematic shift seen in the current \UCN\ lifetime experiment~\cite{Seestrom2014} as well as validating heat flush purification methods.  

We acknowledge the support of the NIST, US Department of Commerce, in providing support for facilities used in this work.  This work is also 
supported in part by the US National Science Foundation under Grant
No.\ PHY-0855593 and by the U.S. Department of Energy, Office of Nuclear Physics, under Contract No. DE-AC02-06CH11357.  This research used resources of ANL's ATLAS facility, which is a DOE Office of Science User Facility.

\bibliographystyle{apsrev4-1}
\bibliography{Helium_PRC.bib}     

\begin{thebibliography}{25}
\bibitem [{dis()}]{disclaimer}%
  \BibitemOpen
  \href@noop {} {}\bibinfo {note} {Certain trade names and company products are
  mentioned in the text or identified in illustrations in order to adequately
  specify the experimental procedure and equipment used. In no case does such
  identification imply recommendation or endorsement by the National Institute
  of Standards and Technology, nor does it imply that the products are
  necessarily the best available for the purpose.}\BibitemShut {Stop}%
\end{thebibliography}

\end{document}